\RequirePackage{lineno}
\documentclass[aps,prd,twocolumn,superscriptaddress,preprintnumbers,amsmath,amssymb,floatfix,raggedbottom]{revtex4}

\usepackage{subfigure}
\usepackage{setspace}
\usepackage{graphicx} % Include figure files
\usepackage{dcolumn}  % Align table columns on decimal point
\usepackage{nccmath}  % Align table columns on decimal point

\graphicspath{{ps}}

\usepackage{xcolor}
\begin{document}

%\vspace*{-3\baselineskip}
%\resizebox{!}{3cm}{\includegraphics{belle.eps}}

%\preprint{\vbox{ \hbox{   }
%Belle Preprint 2021-14
%  KEK Preprint 2021-11
%                        \hbox{Publication 590 }
%                        \hbox{Belle Preprint 2021-14}
%                        \hbox{KEK Preprint 2021-11}
%                        \hbox{BN1554 Final Review Version, June 2021}
%                        \hbox{Intended for {\it Physical Review D}}
%                        \hbox{Author: John Yelton}
  		              % \hbox{hep-ex nnnn}
%}}

%\bigskip

\title{ \quad\\[1.5cm]Measurement of the Masses and Widths of the $\Sigma_c(2455)^+$ and $\Sigma_c(2520)^+$ Baryons}

%%%% >>>>> insert the authorlist here. BEFORE the abstract !!!!! <<<<<
%%%% >>>>> from the authorship confirmation web page
%%% Name the file author.tex and use \input{author} to insert into your latex file.

%%% Paper:    Radiative decays of orbitally excited Xi_c baryons
%%% Journal:  Physical Review D Rapid Communications
%%% Contacts: J. Yelton (yelton@phys.ufl.edu)
%%% Non-responding authors or those who said NO are commented out.
%%% ====================================================================
%%% Click the RELOAD button on your web browser to see the updated file.
%%% ====================================================================
%%% Use \input{author} to insert this material into your latex file.
%%%%% Force institutions to appear in alphabetical order when typeset.

%%% Paper:    Sigma_c(2455)+ and Sigma_c(2520)+
%%% Journal:  Physical Review D
%%% Contacts: J. Yelton (yelton@phys.ufl.edu)
%%% Non-responding authors or those who said NO are commented out.
%%% ====================================================================
%%% Click the RELOAD button on your web browser to see the updated file.
%%% ====================================================================
%%% Use \input{author} to insert this material into your latex file.
%%%%% Force institutions to appear in alphabetical order when typeset.

\noaffiliation
\affiliation{Department of Physics, University of the Basque Country UPV/EHU, 48080 Bilbao}
\affiliation{University of Bonn, 53115 Bonn}
\affiliation{Brookhaven National Laboratory, Upton, New York 11973}
\affiliation{Budker Institute of Nuclear Physics SB RAS, Novosibirsk 630090}
\affiliation{Faculty of Mathematics and Physics, Charles University, 121 16 Prague}
\affiliation{Chonnam National University, Gwangju 61186}
\affiliation{University of Cincinnati, Cincinnati, Ohio 45221}
\affiliation{Deutsches Electronen-Synchrotron, 22607 Hamburg}
\affiliation{University of Florida, Gainesville, Florida 32611}
\affiliation{Department of Physics, Fu Jen Catholic University, Taipei 24205}
\affiliation{Key Laboratory of Nuclear Physics and Ion-beam Application (MOE) and Institute of Modern Physics, Fudan University, Shanghai 200443}
\affiliation{Justus-Liebig-Universit\"at Gie\ss{}en, 35392 Gie\ss{}en}
\affiliation{Gifu University, Gifu 501-1193}
\affiliation{SOKENDAI (The Graduate University for Advanced Studies), Hayama 240-0193}
\affiliation{Gyeongsang National University, Jinju 52828}
\affiliation{Department of Physics and Institute of Natural Sciences, Hanyang University, Seoul 04763}
\affiliation{University of Hawaii, Honolulu, Hawaii 96822}
\affiliation{High Energy Accelerator Research Organization (KEK), Tsukuba 305-0801}
\affiliation{National Research University Higher School of Economics, Moscow 101000}
\affiliation{Forschungszentrum J\"{u}lich, 52425 J\"{u}lich}
\affiliation{IKERBASQUE, Basque Foundation for Science, 48013 Bilbao}
\affiliation{Indian Institute of Science Education and Research Mohali, SAS Nagar, 140306}
\affiliation{Indian Institute of Technology Bhubaneswar, Satya Nagar 751007}
\affiliation{Indian Institute of Technology Guwahati, Assam 781039}
\affiliation{Indian Institute of Technology Hyderabad, Telangana 502285}
\affiliation{Indian Institute of Technology Madras, Chennai 600036}
\affiliation{Indiana University, Bloomington, Indiana 47408}
\affiliation{Institute of High Energy Physics, Chinese Academy of Sciences, Beijing 100049}
\affiliation{Institute of High Energy Physics, Vienna 1050}
\affiliation{Institute for High Energy Physics, Protvino 142281}
\affiliation{INFN - Sezione di Napoli, I-80126 Napoli}
\affiliation{INFN - Sezione di Roma Tre, I-00146 Roma}
\affiliation{INFN - Sezione di Torino, I-10125 Torino}
\affiliation{Advanced Science Research Center, Japan Atomic Energy Agency, Naka 319-1195}
\affiliation{J. Stefan Institute, 1000 Ljubljana}
\affiliation{Institut f\"ur Experimentelle Teilchenphysik, Karlsruher Institut f\"ur Technologie, 76131 Karlsruhe}
\affiliation{Kennesaw State University, Kennesaw, Georgia 30144}
\affiliation{Department of Physics, Faculty of Science, King Abdulaziz University, Jeddah 21589}
\affiliation{Kitasato University, Sagamihara 252-0373}
\affiliation{Korea Institute of Science and Technology Information, Daejeon 34141}
\affiliation{Korea University, Seoul 02841}
\affiliation{Kyoto Sangyo University, Kyoto 603-8555}
\affiliation{Kyungpook National University, Daegu 41566}
\affiliation{Universit\'{e} Paris-Saclay, CNRS/IN2P3, IJCLab, 91405 Orsay}
\affiliation{P.N. Lebedev Physical Institute of the Russian Academy of Sciences, Moscow 119991}
\affiliation{Faculty of Mathematics and Physics, University of Ljubljana, 1000 Ljubljana}
\affiliation{Ludwig Maximilians University, 80539 Munich}
\affiliation{Luther College, Decorah, Iowa 52101}
\affiliation{Malaviya National Institute of Technology Jaipur, Jaipur 302017}
\affiliation{Faculty of Chemistry and Chemical Engineering, University of Maribor, 2000 Maribor}
\affiliation{Max-Planck-Institut f\"ur Physik, 80805 M\"unchen}
\affiliation{School of Physics, University of Melbourne, Victoria 3010}
\affiliation{University of Mississippi, University, Mississippi 38677}
\affiliation{University of Miyazaki, Miyazaki 889-2192}
\affiliation{Moscow Physical Engineering Institute, Moscow 115409}
\affiliation{Graduate School of Science, Nagoya University, Nagoya 464-8602}
\affiliation{Universit\`{a} di Napoli Federico II, I-80126 Napoli}
\affiliation{Nara Women's University, Nara 630-8506}
\affiliation{National Central University, Chung-li 32054}
\affiliation{National United University, Miao Li 36003}
\affiliation{Department of Physics, National Taiwan University, Taipei 10617}
\affiliation{H. Niewodniczanski Institute of Nuclear Physics, Krakow 31-342}
\affiliation{Nippon Dental University, Niigata 951-8580}
\affiliation{Niigata University, Niigata 950-2181}
\affiliation{Novosibirsk State University, Novosibirsk 630090}
\affiliation{Osaka City University, Osaka 558-8585}
\affiliation{Pacific Northwest National Laboratory, Richland, Washington 99352}
\affiliation{Panjab University, Chandigarh 160014}
\affiliation{Peking University, Beijing 100871}
\affiliation{University of Pittsburgh, Pittsburgh, Pennsylvania 15260}
\affiliation{Research Center for Nuclear Physics, Osaka University, Osaka 567-0047}
\affiliation{Dipartimento di Matematica e Fisica, Universit\`{a} di Roma Tre, I-00146 Roma}
\affiliation{Department of Modern Physics and State Key Laboratory of Particle Detection and Electronics, University of Science and Technology of China, Hefei 230026}
\affiliation{Seoul National University, Seoul 08826}
\affiliation{Soongsil University, Seoul 06978}
\affiliation{Sungkyunkwan University, Suwon 16419}
\affiliation{School of Physics, University of Sydney, New South Wales 2006}
\affiliation{Department of Physics, Faculty of Science, University of Tabuk, Tabuk 71451}
\affiliation{Tata Institute of Fundamental Research, Mumbai 400005}
\affiliation{Department of Physics, Technische Universit\"at M\"unchen, 85748 Garching}
\affiliation{School of Physics and Astronomy, Tel Aviv University, Tel Aviv 69978}
\affiliation{Toho University, Funabashi 274-8510}
\affiliation{Department of Physics, Tohoku University, Sendai 980-8578}
\affiliation{Earthquake Research Institute, University of Tokyo, Tokyo 113-0032}
\affiliation{Department of Physics, University of Tokyo, Tokyo 113-0033}
\affiliation{Tokyo Institute of Technology, Tokyo 152-8550}
\affiliation{Tokyo Metropolitan University, Tokyo 192-0397}
\affiliation{Virginia Polytechnic Institute and State University, Blacksburg, Virginia 24061}
\affiliation{Wayne State University, Detroit, Michigan 48202}
\affiliation{Yamagata University, Yamagata 990-8560}
\affiliation{Yonsei University, Seoul 03722}
  \author{J.~Yelton}\affiliation{University of Florida, Gainesville, Florida 32611} % Florida
  \author{I.~Adachi}\affiliation{High Energy Accelerator Research Organization (KEK), Tsukuba 305-0801}\affiliation{SOKENDAI (The Graduate University for Advanced Studies), Hayama 240-0193} % KEK
  \author{J.~K.~Ahn}\affiliation{Korea University, Seoul 02841} % Korea
  \author{H.~Aihara}\affiliation{Department of Physics, University of Tokyo, Tokyo 113-0033} % Tokyo
  \author{S.~Al~Said}\affiliation{Department of Physics, Faculty of Science, University of Tabuk, Tabuk 71451}\affiliation{Department of Physics, Faculty of Science, King Abdulaziz University, Jeddah 21589} % Tabuk
  \author{D.~M.~Asner}\affiliation{Brookhaven National Laboratory, Upton, New York 11973} % BNL
  \author{H.~Atmacan}\affiliation{University of Cincinnati, Cincinnati, Ohio 45221} % Cincinnati
  \author{V.~Aulchenko}\affiliation{Budker Institute of Nuclear Physics SB RAS, Novosibirsk 630090}\affiliation{Novosibirsk State University, Novosibirsk 630090} % BINP
  \author{T.~Aushev}\affiliation{National Research University Higher School of Economics, Moscow 101000} % HSE
  \author{R.~Ayad}\affiliation{Department of Physics, Faculty of Science, University of Tabuk, Tabuk 71451} % Tabuk
% ...
% Lebedev
\collaboration{The Belle Collaboration}
\noaffiliation

%% end author list

\begin{abstract}
Using 
980 ${\rm fb}^{-1}$ of data {collected} with the Belle detector
operating at the KEKB asymmetric-energy $e^+e^-$ collider, we report the measurements
of the masses, and the first measurements of the instrinsic widths, of the
$\Sigma_c(2455)^+$ and $\Sigma_c(2520)^+$ charmed baryons.
We find 
$M(\Sigma_c(2455)^+)-M(\Lambda_c^+) = 166.17\pm 0.05^{+0.16}_{-0.07}\ {\rm MeV}/c^2$, 
$\Gamma(\Sigma_c(2455)^+) = 2.3 \pm 0.3 \pm 0.3\ {\rm MeV/c^2}$,
$M(\Sigma_c(2520)^+)-M(\Lambda_c^+) = 230.9 \pm 0.5 ^{+0.5}_{-0.1}\ {\rm MeV}/c^2$, 
and $\Gamma(\Sigma_c(2520)^+) = 17.2^{+2.3\ +3.1}_{-2.1\ -0.7}\ {\rm MeV}/c^2$,
where the uncertainties are statistical and systematic, respectively.
These measurements can be used to test models of the underlying quark structure
of the $\Sigma_c$ states.

\end{abstract}

%\pacs{14.20.Lq}

\maketitle

%%%% >>>> keep the final version single-spaced

{\renewcommand{\thefootnote}{\fnsymbol{footnote}}}
\setcounter{footnote}{0}
\raggedbottom
\section*{Introduction}

The $\Sigma_c$ charmed baryons consist of a charm quark in combination with a spin-1 light
($uu$, $ud$ or $dd$) diquark. 
The lowest of these states are the $\Sigma_c(2455)$ isotriplet which have $J^P=\frac{1}{2}^+$, with the next most massive
being the $J^P=\frac{3}{2}^+$ $\Sigma_c(2520)$ isotriplet. 
All these six states decay strongly to $\Lambda_c^+\pi$.
The doubly-charged and neutral states, both of which decay with the emission of a charged pion, have been well studied.
The most precise measurements of their masses and widths have been made~\cite{SIGC} by the Belle Collaboration using
the same dataset as the analysis presented here. 
However, $\pi^0$ transitions have lower efficiency, higher backgrounds and
inferior resolution to $\pi^{\pm}$ transitions, so there is comparatively little 
experimental information on the singly-charged $\Sigma_c^{+}$ states~\cite{CC}.

All mass measurements of the $\Sigma_c$ baryons have been made with respect to the $\Lambda_c^+$ mass, as the 
resolution of these mass differences (denoted $\Delta(M)$) is superior to that of the individual baryons. 
The CLEO Collaboration has measured $\Delta(M)$ for both the $\Sigma_c(2455)^+$
and $\Sigma_c(2520)^+$ states~\cite{CLEO}, but were only able to set limits on their intrinsic widths. The large Belle dataset
allows for much more precise measurements of the masses of these
particles than has been possible hitherto, and also the first measurements of their widths. 

Measurements of the masses of all members of the two isotriplets allow tests of models of isospin
mass splittings. In the model of Yang and Kim~\cite{YK}, for instance, the mass splittings from the following four sources add:
the electromagnetic corrections due to the light quarks, the differences of the masses
of the $u$ and $d$ quarks, the hyperfine interactions between the light quarks, and the Coulomb interactions between 
the soliton and charm quark. 
Most mass models predict that the singly-charged states should have
masses a little lower than their doubly-charged and neutral analogs~\cite{theories} and this is true in the 
limited precision measurements made to date~\cite{PDG}.

The natural width of the $\Sigma_c(2455)^+$ is predicted to 
be somewhat larger than its isospin partners; this is mostly because of the effect of the $\pi^{\pm}/\pi^0$ mass difference 
on the available phase space
for the decay. There is also a possibility that electromagnetic decays are non-negligible. 
Cheng and Chua~\cite{Cheng} predict $\Gamma(\Sigma_c(2455)^+)=2.3^{+0.1}_{-0.2}\ {\rm MeV}/c^2$ using the experimental 
value of $\Gamma(\Sigma_c(2455)^{++})=1.94^{+0.08}_{-0.16}\ {\rm MeV}/c^2$ as input.
For the $\Sigma_c(2520)$,
it is expected that the instrinsic width of the $\Sigma_c^{+}$ will be similar to those
measured for its isospin partners of 
$\Gamma(\Sigma_c(2520)^{++})=14.8^{+0.3}_{-0.4}\ {\rm MeV}/c^2$ and
$\Gamma(\Sigma_c(2520)^{0})=15.3^{+0.4}_{-0.5}\ {\rm MeV}/c^2$, 
respectively; the two effects listed above are expected to be small compared with these values.

In addition to checking the quark model predictions, the parameters of the $\Sigma_c(2455)^+$ 
are vital in studies of the
$\Lambda_c(2595)^+$, whose pole mass appears to be between the $\Sigma_c(2455)^+\pi^0$ and $\Sigma_c(2455)^{++}\pi^-$ thresholds. 
This particle, although 
generally considered to be an orbitally excited heavy-quark light-diquark state, 
has been conjectured to have different underlying quark structure~\cite{2593}.
The threshold behavior, and thus measurement of the pole mass and width, 
of the $\Lambda_c(2595)^+$ is critically dependent on the masses and widths of the $\Sigma_c$ particles.   

\begin{table*}[htb]
\caption{Parameters of the double-Gaussian resolution function derived from the Monte Carlo program.}
\begin{spacing}{1.0}
\begin{tabular}
 { c | c | c |c |c}

\hline 
Particle  & $\sigma_{\rm narrow}\ ({\rm MeV}/c^2)$   & $\sigma_{\rm wide}\ ({\rm MeV}/c^2)$ &   ${\rm Area}_{\rm wide}/{\rm Area}_{\rm narrow}$ & Mass Offset $({\rm MeV}/c^2)$ \\

\hline
$\Sigma_c(2455)^+$ & 1.473             &   2.932         &  0.97       &  0.078       \\
$\Sigma_c(2520)^+$ & 2.23             &  4.22          &  3.06       &     0.08      \\

\hline
\hline
\end{tabular}
\end{spacing}
\label{tab:resol}
\end{table*}

\section*{Detector and Dataset}

This analysis uses a data sample of $e^+e^-$ annihilations recorded by the Belle detector~\cite{Belle} 
operating at the KEKB asymmetric-energy $e^+e^-$
collider~\cite{KEKB}. It corresponds to an integrated luminosity of 980 ${\rm fb}^{-1}$.
The majority of these data were taken with the accelerator energy tuned for production of the $\Upsilon(4S)$ resonance, 
as this is optimum for investigation of $B$ decays.
However, the $\Sigma_c$ particles in this analysis are produced in continuum charm production and are of
higher momentum than those that are decay products of $B$ mesons. 
This allows the use of the complete Belle data set which includes data taken at beam energies 
corresponding to the other $\Upsilon$ resonances and the nearby continuum.
The Belle detector is a large solid-angle spectrometer comprising six sub-detectors: 
the Silicon Vertex Detector (SVD), the 50-layer Central
Drift Chamber (CDC), the Aerogel Cherenkov Counter (ACC), the Time-of-Flight scintillation counter (TOF),
the electromagnetic calorimeter (ECL), and the
$K_L$ and muon detector. A superconducting solenoid produces a 1.5 T magnetic field throughout 
the first five of these sub-detectors.
The detector is described in detail elsewhere~\cite{Belle}. 
Two inner detector configurations were used. The first comprised a 2.0 cm radius
 beampipe and a 3-layer silicon vertex detector, and the second a 1.5 cm radius beampipe 
and a 4-layer silicon detector and a small-cell 
inner drift chamber.

\section*{Analysis}

We study $\Sigma_c^+$ baryons from the decay chain $\Sigma_c^+\to\Lambda_c^+ \pi^0,\Lambda_c^+ \to p K^- \pi^+$.
The decays are reconstructed from combinations of charged particles
measured using the tracking system, and neutral particles measured in the ECL. 
Final-state charged particles, $\pi^{+}, K^{-}$, and $p$, are selected using the likelihood
information from the tracking (SVD, CDC) and charged-hadron identification (CDC, ACC, TOF) systems into a combined likelihood,
${\cal L}(h1:h2) = {\cal L}_{h1}/({\cal L}_{h1} + {\cal L}_{h2})$
where $h1$ and $h2$ are $p$, $K$, and $\pi$ as appropriate~\cite{Nakano}. We require proton candidates
to have ${\cal L}(p:K)>0.6$ and ${\cal L}(p:\pi)>0.6$,
kaon candidates to have ${\cal L}(K:p)>0.6$ and ${\cal L}(K:\pi)>0.6$,
and pions to have requirements of
${\cal L}(\pi:K)>0.6$ and ${\cal L}(\pi:p)>0.6$. 
The efficiencies of thse hadron identification requirements are about 90\%, 90\%, and 93\% 
for pions, kaons and protons, respectively. The probability to misidentify a pion (kaon) track as a kaon (pion)
track is about 9 (10)\%, and the momentum averaged
probability to misidentify a pion or kaon track as a proton track is about 5\%.
Combinations of $pK^-\pi^+$ candidates with an invariant mass
within 3.9 ${\rm MeV}/c^2$ 
(approximately two standard deviations ($\sigma$)) of the $\Lambda_c^+$ were retained as $\Lambda_c^+$ candidates.
The number of events having more than one $\Lambda_c^+$ candidate which share a daughter particle is approximately 1\%.

The $\pi^0$ candidates are reconstructed from two detected neutral clusters in the ECL each consistent with being
due to a photon and each with an energy greater than 50 ${\rm MeV}$ in the laboratory frame. 
The invariant mass of the photon pair is required to be
within $5.4\ {\rm MeV}/c^2$\ ($\approx\ 2\sigma$) of the nominal $\pi^0$ mass. 
The two photons are then constrained to this mass to improve the momentum resolution of the $\pi^0$.

To optimize the requirements specific to this analysis, a simulated data set is constructed using
a combination of the decays under study and $e^+e^-$ hadronic events generated by PYTHIA~\cite{PYTHIA}.
We find that the following requirements are optimal for the highest $\Sigma_c(2455)^+$ 
signal significance: the momentum of the $\Sigma_c^+$ candidate in the $e^+e^-$ center-of-mass 
frame, $p^*\ >\ 2.6\ {\rm GeV}/c$; the momentum of the $\pi^0$ in laboratory frame, $p\ >\ 200\ {\rm MeV}/c$.
%We find that requirements that the center-of-mass
%momentum of the $\Sigma_c^+$ candidate have $p^* > 2.6\ {\rm GeV}/c$ and the 
%laboratory momentum of the $\pi^0$ be $p > 200\ {\rm MeV}/c$
%are optimal for best $\Sigma_c(2455)$ signal significance. 

The Monte Carlo (MC) simulation is performed using a 
GEANT-based MC simulation~\cite{Geant} to model the
response of the detector. The photon energy response in the simulation is corrected 
to take into account the data-MC difference of resolution
based on studies of mass resolution in the
decays $\pi^0\to\gamma\gamma$, $\eta\to\gamma\gamma$, and $D^{*0}\to D^0\gamma$~\cite{Umberto, Mizuk}.
The resolution of the $\Sigma_c(2455,2520)^+$ mass 
peaks is parameterized by double-Gaussian resolution functions with a small offset in the peak mass allowed.
The parameters of these functions are shown in Table~\ref{tab:resol} and the statistical uncertainties in these
values are negligible. 
It is immediately clear that knowledge of the $\Sigma_c(2455)^+$ signal resolution is vital 
as it is similar to the expected intrinsic width. 
To further check the MC simulation, a study was made of the decay 
$D^{*+}\to D^+ \pi^0$, where $D^+\to K^-\pi^+\pi^+$.  
This decay has almost the same final state as the one under consideration, 
similar momentum distribution, is much more copiously produced, 
and has a very small and well-known intrinsic width. 
The resolution of this mode is found to be 3\% larger in data than in MC simulations,
and the reconstructed mass was found to be $0.020\pm0.015\ {\rm MeV}/c^2$ lower,
where the uncertainty is due to the Particle Data Group value~\cite{PDG} as our statistical uncertainties are negligible. 
We take these comparisons into account in the considerations of the systematic uncertainties of our  $\Sigma_c^+$ measurements.

\begin{figure*}[htb]
\includegraphics[width=7.0in]{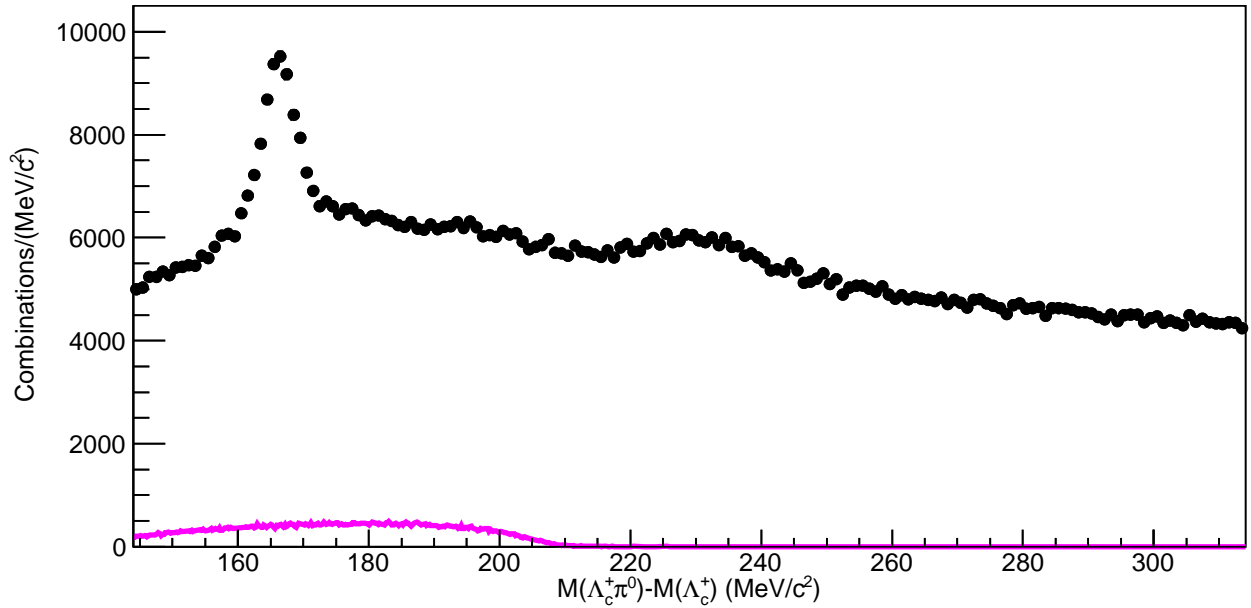}
\caption{The mass difference $M(\Lambda_c^+\pi^0)-M(\Lambda_c^+)$ for the entire mass range of interest. The line shows the expected
contribution from $\Lambda_c(2625)^+\to\Lambda_c^+\pi^0\pi^0$ decays assuming uniform 3-body phase-space. }
\label{fig:Figure1}
\end{figure*}

\begin{figure}[htb]
\includegraphics[width=3.1in]{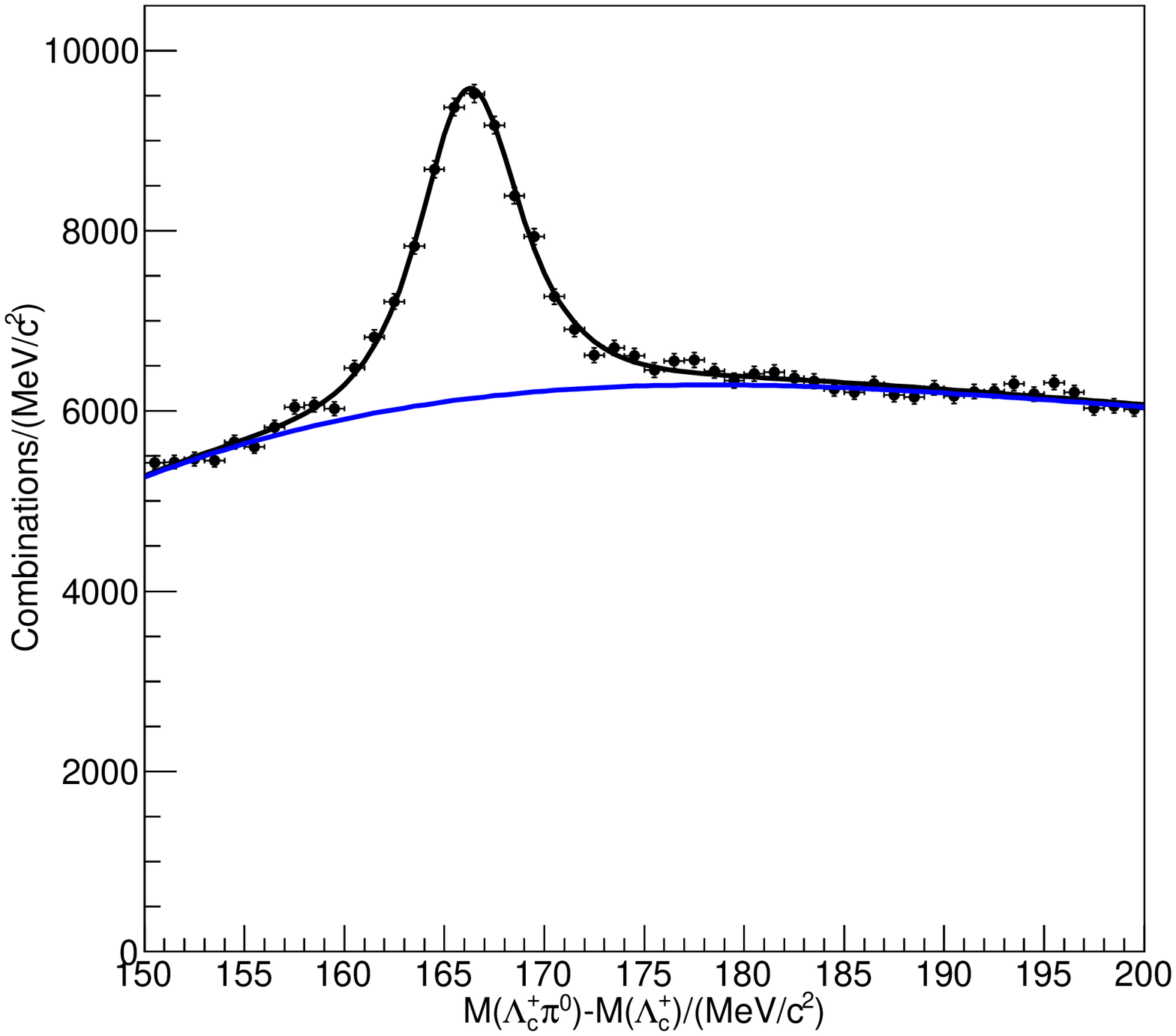}

\caption{The mass difference $M(\Lambda_c^+\pi^0)-M(\Lambda_c^+)$ in the region of the $\Sigma_c(2455)^+$. The 
fit to the data is described in the text, and the lower line shows its contribution from the background function. }
\label{fig:S2455}

\end{figure}

\begin{figure}[htb]
\includegraphics[width=3.1in]{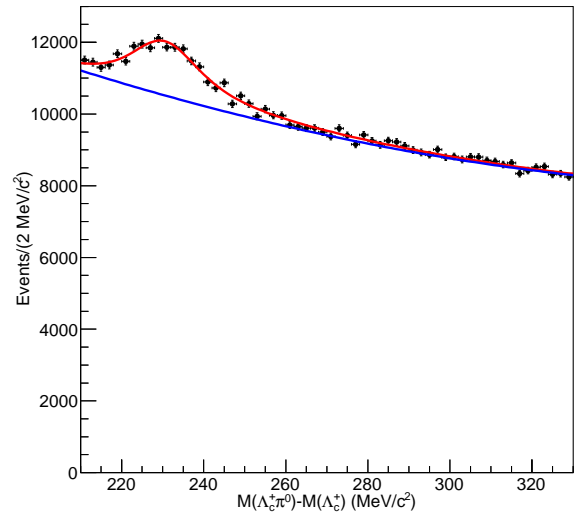}

\caption{The mass difference $M(\Lambda_c^+\pi^0)-M(\Lambda_c^+)$ in the region of the $\Sigma_c(2520)^+$. The 
fit to the data is described in the text, and the lower line shows its contribution from the background function. }
\label{fig:S2520}

\end{figure}

Figure~\ref{fig:Figure1} shows the invariant mass distribution for the $\Lambda_c^+\pi^0$ candidates. 
Clear signals
are seen corresponding to $\Sigma_c(2455)^+$ and $\Sigma_c(2520)^+$ production. In addition, we see the large
enhancement up to a mass difference of $\approx 200\ {\rm MeV}/c^2$ due to $\Lambda_c(2625)^+\to\Lambda_c^+\pi^0\pi^0$ 
decays as they produce $\Lambda_c^+\pi^0$ combinations with mass differences up to the kinematic limit of 
$M(\Lambda_c^+(2625))-M(\Lambda_c^+)-M(\pi^0) = 207\ {\rm MeV}/c^2$. The simulated shape of this component is
shown in Fig.\ref{fig:Figure1}. 
There is also a possible enhancement due to $\Lambda_c(2625)^+\to\Sigma_c^+\pi^0$ decays
that may produce an enhancement at around $194\ {\rm MeV}/c^2$. These enhancements were anticipated because of isospin 
symmetry, but these particular decays have never been studied. The ``cusp'' behavior at around $200\ {\rm MeV/c^2}$
is particularly problematic as its shape depends on the relative contributions of 3-body decays, decays proceeding 
through virtual $\Sigma_c(2520)$ production, and the interference between these two~\cite{Arifi}. Rather than 
fitting the entire spectrum, we decided to find the signal parameters from fits performed to limited-range subsets of this data
which do not include this cusp region. 
These are shown in Fig.~\ref{fig:S2455} and Fig.~\ref{fig:S2520} for the $\Sigma_c(2455)^+$ and $\Sigma_c(2520)^+$ regions,
respectively. The results of a 
global fit to Fig.~\ref{fig:Figure1} will be taken into account in the systematic uncertainty determination.

A fit is made to Fig.~\ref{fig:S2455} using a third-order Chebychev polynomial function to represent the background, and a P-wave
relativistic Breit-Wigner function convolved with the previously described double-Gaussian resolution function, 
taking into account the small
mass offset. The Breit-Wigner signal function includes a Blatt-Weisskopf barrier factor~\cite{BW},
with radius parameter of $R = 3~{\rm GeV}^{-1}$~\cite{BW2}. 
The results of this fit are a mass difference of 
$\Delta(M) = 166.17\pm0.05\ {\rm MeV}/c^2$ and 
$\Gamma = 2.3\pm0.3\ {\rm MeV}/c^2$. 
The fit is made using a maximum-likelihood method to a large number of small bins so that any uncertainty due to bin size is negligible.
A convenient test of the goodness-of-fit is the $\chi^2$ per degree of freedom (reduced $\chi^2$)
for the distribution as shown,
and for Fig.~\ref{fig:S2455} this reduced $\chi^2$ is $49.2/43 = 1.14$.

A fit is made to Fig.~\ref{fig:S2520} using a second-order Chebychev polynomial function to represent the background, 
and a P-wave relativistic Breit-Wigner function convolved with the double-Gaussian resolution function described above. 
The results of this fit are a mass difference of 
$\Delta(M) = 230.9\pm0.5\ {\rm MeV}/c^2$ 
and $\Gamma = 17.2^{+2.3}_{-2.1}\ {\rm MeV}/c^2$.
The reduced $\chi^2$ of the fit for the plot as shown is $52.8/44 =1.20$.

\section*{Systematic uncertainties}

In all four measurements, we take the systematic uncertainty due to fitting as the maximum variation of the measured parameters 
using different fitting functions which produce acceptable fits to the data. 
For the $\Sigma_c(2455)^+$ we vary the power of the polynomial background function from 2 to 4, allow the possibility of a satellite
peak due to $\Lambda_c(2625)^+\to\Sigma_c^+\pi^0$ decays as such decays are expected at a low level, investigate the changes in 
parameters with small changes to the fitting ranges, 
and also compare the results of the fit shown in Fig.~\ref{fig:S2455} with the results of a global fit to Fig.~\ref{fig:Figure1} which
includes contributions from $\Lambda_c(2595)^+$ and $\Lambda_c(2625)^+$ decays.

The differences in the values obtained using reasonable variations of the Blatt-Weisskopf 
barrier parameter, $R$, were found to be small.
As the measurement of the $D^{*+}$ width 
indicates a possible underestimation of the detector resolution by 3\%, we also perform a fit using a resolution 6\% higher than that 
found from MC simulations and conservatively take the change in the parameters as the systemetic uncertainty arising from the uncertainty
in the mass resolution. We similarly study the variation of the $\Sigma_c(2520)^+$ measured parameters to estimate the associated
systematic uncertainties, but here we cannot
reduce the order of the polynomial background function as a first order polynomial does not produce a satistfactory fit to the
data. 

For the systematic uncertainty due to the energy scale, we allow for the possibility that the $D^{*+}$ mass is measured up to
$0.035\ {\rm MeV}/c^2$ lower than the true mass. We make the conservative assumption that the difference between the measured
and canonical masses of the $D^{*+}$ is entirely due to a miscalibration of our
photon energy scale, and use MC simulation to estimate how a change in the $D^{*+}$ mass, which is a decay with a 
very small four-momentum-squared $(q^2)$ associated with it, translates to a change in mass for a particle decaying with a larger $q^2$. 
The result of this study is a 
possible upward shift of $0.15\ {\rm MeV}/c^2$ in the mass of the $\Sigma_c(2455)^+$
and $0.3\ {\rm MeV}/c^2$ for the $\Sigma_c(2520)^+$. The systematic uncertainty estimations are tabulated in Table~\ref{tab:syst}.

\begin{table}[htb]
\caption{Contributions to the systematic uncertainties of the mass difference and width measurements of the two states in ${\rm MeV}/c^2$.}
\begin{spacing}{1.2}
\begin{tabular}
 { c | c |c |c|c}

\hline 
  & \multicolumn{2}{c|} {$\Sigma_c(2455)^+$ }  & \multicolumn{2}{c} {$\Sigma_c(2520)^+$} \\
\hline
  & $\Delta(M)$   & $\Gamma$       & $\Delta(M)$     & $\Gamma$ \\
 
\hline
Background Function & $^{+0.00}_{-0.07}$         &  $^{+0.30}_{-0.04}$  & $^{+0.4}_{-0.0}$ & $^{+3.0}_{-0.5}$ \\
Signal Function     & $^{+0.06}_{-0.01}$         &  $^{+0.01}_{-0.33}$  & $^{+0.0}_{-0.1}$ & $^{+0.7}_{-0.5}$ \\
Photon Energy Scale         & $^{+0.15}_{-0.00}$ &  $^{+0.00}_{-0.00}$  & $^{+0.3}_{-0.0}$ & $^{+0.0}_{-0.0}$ \\

\hline
\hline
\end{tabular}
\end{spacing}
\label{tab:syst}
\end{table}
\raggedbottom
\section*{Discussion}

The measured instrinsic widths are consistent with the quark model predictions~\cite{Cheng},
namely that it is the same as the widths of their isospin partners, except for a small change due to the increased
phase-space available.  The measured mass differences are consistent with, but more
precise than, the previous measurements~\cite{PDG} and confirm the picture in which the singly-charged
states are slightly lower in mass than their isospin partners. According to the model first proposed
by Franklin~\cite{Franklin} the value of the mass relationship
$M(\Sigma_c^{++})+M(\Sigma_c^0)-2\times M(\Sigma_c^+)$ should be the same for the $\Sigma_c(2455)$ 
and $\Sigma_c(2520)$ isotriplets.
Combining our measurements for the singly-charged $\Sigma_c$ states with those of the Particle Data Group~\cite{PDG}
for the others, we find values of $2.46^{+0.17}_{-0.34}$ and $2.2^{+1.0}_{-1.4}\ {\rm MeV}/c^2$ for the two systems,
respectively, consistent with the model.
Yang and Kim~\cite{YK} further predict the mass difference between the singly-charged and neutral
$\Sigma_c$ baryons should be the same as those between the analogous $\Xi_c^{\prime}$ and $\Xi_c^*$ states,
and our results are also consistent with this prediction.

\section*{Conclusions}

We measure the mass difference of the $\Sigma_c(2455)^+$ with respect to the $\Lambda_c^+$ to be
$\Delta(M) = 166.17\pm0.05^{+0.16}_{-0.07}\ {\rm MeV}/c^2$ and its instrinsic width
$\Gamma = 2.3\pm0.3\pm0.3\ {\rm MeV}/c^2$. For the $\Sigma_c(2520)^+$ the analogous values are
$\Delta(M) = 230.9\pm0.5^{+0.5}_{-0.1}\ {\rm MeV}/c^2$ 
and $\Gamma = 17.2^{+2.3 +3.1}_{-2.1 -0.7}\ {\rm MeV}/c^2$.
These are the first non-zero measurements of the instrinsic widths of these particles and 
show no deviation from the expectations based upon 
the precise measurements of their isospin partners made using the standard
quark model.

\section*{Acknowledgments}

%----------- Long version, for most papers ----------- 
%----------- Long version, for most papers ----------- 
We thank the KEKB group for the excellent operation of the
accelerator; the KEK cryogenics group for the efficient
operation of the solenoid; and the KEK computer group, and the Pacific Northwest National
Laboratory (PNNL) Environmental Molecular Sciences Laboratory (EMSL)
computing group for strong computing support; and the National
Institute of Informatics, and Science Information NETwork 5 (SINET5) for
valuable network support.  We acknowledge support from
the Ministry of Education, Culture, Sports, Science, and
Technology (MEXT) of Japan, the Japan Society for the 
Promotion of Science (JSPS), and the Tau-Lepton Physics 
Research Center of Nagoya University; 
the Australian Research Council including grants
DP180102629, % Sevior
DP170102389, % Varvell
DP170102204, % Yabsley
DP150103061, % Urquijo
FT130100303; % Urquijo;
Austrian Federal Ministry of Education, Science and Research (FWF) and
FWF Austrian Science Fund No.~P~31361-N36;
the National Natural Science Foundation of China under Contracts
No.~11435013,  %Zhen-An Liu
No.~11475187,  %Chang-Zheng Yuan
No.~11521505,  %Chang-Zheng Yuan
No.~11575017,  %Cheng-Ping Shen
No.~11675166,  %Wen-Biao Yan
No.~11705209;  %Yi-Ming Li
Key Research Program of Frontier Sciences, Chinese Academy of Sciences (CAS), Grant No.~QYZDJ-SSW-SLH011; % Chang-Zheng Yuan
the  CAS Center for Excellence in Particle Physics (CCEPP); %Chang-Zheng Yuan,
the Shanghai Pujiang Program under Grant No.~18PJ1401000;  %Tao Luo
the Shanghai Science and Technology Committee (STCSM) under Grant No.~19ZR1403000; %Xiaolong Wang
the Ministry of Education, Youth and Sports of the Czech
Republic under Contract No.~LTT17020;
Horizon 2020 ERC Advanced Grant No.~884719 and ERC Starting Grant No.~947006 ``InterLeptons'' (European Union);
the Carl Zeiss Foundation, the Deutsche Forschungsgemeinschaft, the
Excellence Cluster Universe, and the VolkswagenStiftung;
the Department of Atomic Energy (Project Identification No. RTI 4002) and the Department of Science and Technology of India; 
the Istituto Nazionale di Fisica Nucleare of Italy; 
National Research Foundation (NRF) of Korea Grant
Nos.~2016R1\-D1A1B\-01010135, 2016R1\-D1A1B\-02012900, 2018R1\-A2B\-3003643,
2018R1\-A6A1A\-06024970, 2018R1\-D1A1B\-07047294, 2019K1\-A3A7A\-09033840,
2019R1\-I1A3A\-01058933;
Radiation Science Research Institute, Foreign Large-size Research Facility Application Supporting project, the Global Science Experimental Data Hub Center of the Korea Institute of Science and Technology Information and KREONET/GLORIAD;
the Polish Ministry of Science and Higher Education and 
the National Science Center;
the Ministry of Science and Higher Education of the Russian Federation, Agreement 14.W03.31.0026, % from 15.02.2018
and the HSE University Basic Research Program, Moscow; % from 15.04.2021
University of Tabuk research grants
S-1440-0321, S-0256-1438, and S-0280-1439 (Saudi Arabia);
the Slovenian Research Agency Grant Nos. J1-9124 and P1-0135;
Ikerbasque, Basque Foundation for Science, Spain;
the Swiss National Science Foundation; 
the Ministry of Education and the Ministry of Science and Technology of Taiwan;
and the United States Department of Energy and the National Science Foundation.


\begin{thebibliography}{99}
\bibitem{CC}
Throughout this paper, 
the inclusion of the charge-conjugate is implied.
\end{thebibliography}
\end{document}